\documentclass[sigconf]{acmart}

\setcopyright{none}
\renewcommand\footnotetextcopyrightpermission[1]{}
\acmConference[WWW '27]{The Web Conference 2027}{May 10--14, 2027}{Dublin, Ireland}
\acmYear{2027}
\acmISBN{}
\acmDOI{}

\usepackage{algorithm}
\usepackage{algorithmic}
\usepackage{booktabs}
\usepackage{multirow}
\usepackage{tabularx}
\usepackage{ragged2e}
\usepackage{microtype}
\usepackage{placeins}
\usepackage{stfloats}

\renewcommand{\arraystretch}{0.95}


\title{Consistency Is the Attack Surface: Detecting Coordinated Retrieval Poisoning with Embedding Geometry}

\author{Zeyan Li}
\affiliation{%
  \institution{Shanghai Jiao Tong University}
  \city{Shanghai}
  \country{China}
}

\author{Qirong Guo}
\affiliation{%
  \institution{The Hong Kong University of Science and Technology (Guangzhou)}
  \city{Guangzhou}
  \country{China}
}

\author{Jianfeng Xu}
\affiliation{%
  \institution{Shanghai Jiao Tong University}
  \city{Shanghai}
  \country{China}
}

\begin{document}

\begin{abstract}
Consistency-based defenses for retrieval-augmented generation (RAG) rest on an assumption that a sparse adversary breaks: a poisoning campaign may modify a negligible fraction of the corpus, yet retrieval concentrates its passages into the reader's context. Across four Natural Questions retrieval seeds, at least one injection reaches the top-$5$ in $94.1\%$ of targeted query--seed rows, and all three form a retrieved majority in $52.9\%$. Under our short-answer protocol, RobustRAG and ReliabilityRAG reach exposure-conditional attack success $0.512$, against $0.576$ for undefended RAG, while lowering clean accuracy from $0.328$ to $0.216$. Coordination, however, leaves a fingerprint: a dense, query-anchored component in the evidence embedding geometry. \textsc{CoordGuard} audits this structure from robustly normalized query-weighted affinities, without poison labels, correct answers, or reader gradients. Its always-filter mode suppresses the selected component whenever one forms; a benign-calibrated gate suppresses it only at unusually high scores and traces a robustness--utility frontier. Pooling four retrieval seeds, e5 always-filtering reduces $\mathrm{ASR}_{\mathrm{exp}}$ to $0.167$ overall and $0.138$ on parametrically unknown queries, at a $13.6$-point clean cost, while a Contriever gate retains clean accuracy $0.315$ at $\mathrm{ASR}_{\mathrm{exp}}=0.331$. The reduction transfers across two readers, three tasks, and multiple retriever--auditor pairs, and count-matched controls attribute it to component-targeted removal rather than deletion volume. These findings recast poison detection from judging truth to auditing the structure of retrieved evidence.
\end{abstract}

\maketitle

\section{Introduction}

\noindent\textbf{Relevance to the Web and to the Security and Privacy track.} Web users increasingly receive answers synthesized by retrieval-augmented AI over Web content, which generative AI now both produces and consumes. We study coordinated poisoning of that retrieval interface and present a training-free, reader-agnostic auditor.

Retrieval-augmented generation (RAG) grounds model outputs in external evidence, but its security depends on the evidence delivered to the reader, not on the corpus in aggregate. A poisoning campaign may modify only a negligible corpus fraction while a retriever concentrates several injected passages into a small top-$k$ context. Corpus-level sparsity therefore does not imply retrieved-set sparsity. Globally sparse corruption can become locally dominant at the reader interface.

\begin{figure}[t]
\centering
\includegraphics[width=0.9\linewidth]{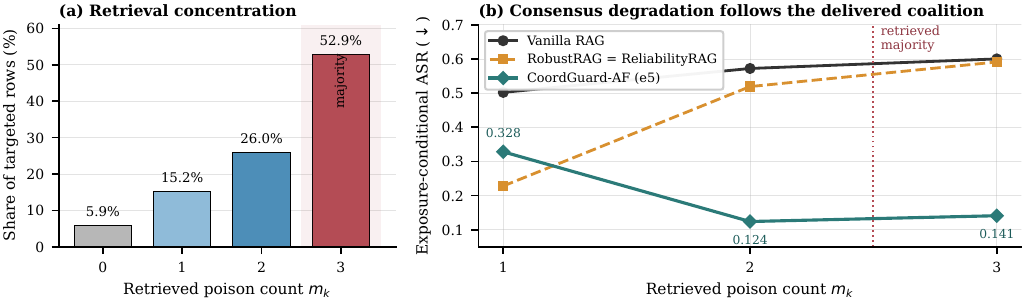}
\caption{Retrieval concentration and consensus degradation on NQ-Qwen (four seeds). (a) Poison reaches the top-$5$ in $94.1\%$ of targeted rows; all three injections form a majority in $52.9\%$. (b) Aggregators degrade as the delivered coalition grows; CoordGuard improves once coordination forms (RobustRAG and ReliabilityRAG coincide).}
\Description{Two-panel figure. The first panel shows the distribution of zero, one, two, or three retrieved poison passages. The second plots exposure-conditional attack success against the retrieved poison count for vanilla RAG, the coincident RobustRAG and ReliabilityRAG curves, and CoordGuard.}
\label{fig:frontier}
\end{figure}

We study \emph{coordinated retrieval poisoning}, in which multiple injected passages use different wording while supporting the same attacker-selected claim. In our headline configuration, up to three injections enter a context of $k{=}5$. Once they become the largest coherent group, and especially when all three form a local majority, agreement among retrieved sources no longer distinguishes reliability from coordination. Majority voting and consistency-graph aggregation can then receive the attacker's mutually supportive coalition as their dominant input even though the corpus modification is small.

This regime creates three challenges. First, consensus can become an attack mechanism instead of a reliability signal. RobustRAG~\cite{roburstrag2024} and ReliabilityRAG~\cite{reliabilityrag2025} provide guarantees under their stated corruption and consistency conditions; our study examines the evidence coalition that remains after retrieval. Second, coordinated poison is a \emph{dense inlier}. Its passages must be relevant enough to enter the top-$k$ and aligned enough to advocate the same target, so relevance pruning, isolation-outlier detection, MMR, and near-duplicate suppression examine the wrong side of the geometry. Third, the same structure can occur benignly when several passages describe the same entity, event, or fact, creating an unavoidable robustness--utility tension for geometry-only intervention.

Figure~\ref{fig:frontier} directly connects retrieval concentration to reader-side failure. On Natural Questions (NQ), retrieval exposes poison in $94.1\%$ of targeted query--seed rows and produces a three-of-five adversarial majority in $52.9\%$. The evaluated consistency aggregators attain exposure-conditional ASR $0.228$, $0.519$, and $0.591$ at realized poison occupancy $m_k{=}1,2,3$; CoordGuard's complementary profile turns favorable only once coordination forms (Section~\ref{sec:main}).

The attack nevertheless tends to leave a structural footprint. Effective poison passages jointly inflate query--passage relevance and passage--passage similarity, producing a dense query-anchored component. We introduce \textsc{CoordGuard}, a training-free, reader-agnostic auditor that constructs candidate components from robustly normalized query-weighted affinities. A continuous score ranks the components and enables either unconditional filtering or a threshold calibrated only on benign retrievals.

Across four NQ retrieval seeds, e5-based \textsc{CoordGuard-AF} reduces $\mathrm{ASR}_{\mathrm{exp}}$ from $0.576$ to $0.167$ overall and to $0.138$ on parametrically unknown queries (clean $0.328\to0.192$ at this endpoint); a Contriever-based gate retains clean accuracy $0.315$ at $\mathrm{ASR}_{\mathrm{exp}}=0.331$. The observed reduction transfers to a second reader, HotpotQA, MS MARCO wrong-claim steering, and multiple retriever--auditor pairs. Count-matched controls show that targeting a query-weighted graph component, instead of merely deleting more passages, is critical, with the selector chosen per operating condition.

\paragraph{Contributions.}
(1) We formalize retrieval concentration through realized poison occupancy and show that reader-side consensus degradation tracks the coalition delivered by retrieval, not corpus corruption rate alone. (2) We introduce \textsc{CoordGuard}, a reader-agnostic query-anchored component audit with unconditional and benign-calibrated operating modes. (3) We evaluate robustness, clean utility, transfer, matched-removal controls, and score-aware adaptation, with clean utility characterized at every operating point. Our setting is inference-time auditing with a fixed retriever and an available auditor embedding; always-filtering is the robustness-oriented endpoint ($\sim90\%$ benign intervention), and deployment uses the benign-calibrated gate.

\section{Related Work}
\label{sec:related}

\paragraph{Retrieval augmentation and corpus poisoning.}
RAG systems~\cite{lewis2020rag} combine a retriever with a parametric reader, using sparse lexical scoring such as BM25~\cite{robertson2009bm25}, dense dual encoders such as DPR~\cite{karpukhin2020dense} and Contriever~\cite{izacard2022contriever}, late-interaction models such as ColBERTv2~\cite{santhanam2022colbertv2}, or fusion-in-decoder architectures~\cite{izacard2021fusion}. Their dependence on an external corpus introduces a distinct attack surface. An adversary can alter what reaches the reader without changing model parameters. PoisonedRAG~\cite{poisonedrag2025} demonstrates targeted knowledge corruption by inserting adversarial documents into the retrieval corpus. We focus on the coordinated case in which several injected passages support one target answer and are jointly retrieved. The central phenomenon is not simply that poison exists in the corpus, but that retrieval concentrates a small campaign into a majority of a small top-$k$ context.

\paragraph{Aggregation and certified retrieval defenses.}
RobustRAG~\cite{roburstrag2024} isolates evidence-conditioned predictions and aggregates them to obtain robustness under bounded retrieval corruption. ReliabilityRAG~\cite{reliabilityrag2025} represents consistency relations among sources and selects reliable evidence through a graph-based rule. These methods establish guarantees under their stated threat and certification conditions; our experiments do not challenge those guarantees. We instead examine an operational regime in which three coordinated corpus injections already occupy a majority of the top-five retrieved passages. At that point, the evaluated majority- and maximum-independent-set rules receive an adversarial coalition as their dominant evidence and may select it. This distinguishes \emph{corpus sparsity} from \emph{retrieved-set sparsity}. A small number of corpus edits need not remain a minority after retrieval. Self-consistency prompting~\cite{wang2022selfconsistency} also exploits repeated agreement, although it is a reasoning-time aggregation method, not a certified retrieval defense.

\paragraph{Context amplification and knowledge conflicts.}
CAD~\cite{shi2024cad}, AdaCAD~\cite{wang2025adacad}, and CoCoA~\cite{wang2025cocoa} balance parametric and contextual distributions during decoding. Their premise is useful when retrieved evidence is more reliable than model memory, but coordinated poisoning reverses that premise by making the context intentionally misleading. More broadly, retrieved evidence can conflict with parametric memory, with other passages, or with entity-specific facts~\cite{xu2024knowledgeconflicts,xie2024adaptive,longpre2021entity,kortukov2024contextmemory,du2024context}. Prior work studies the internal mechanisms of such conflicts~\cite{jin2024cutting}, resolution strategies~\cite{jin2024tugofwar,li2025taming}, truth-aware context selection~\cite{yu2024truth}, robustness to irrelevant context~\cite{yoran2024robust}, and preference-based knowledge selection~\cite{zhang2024knowpo}. Astute RAG~\cite{wang2024astute}, RA-RAG~\cite{hwang2025rarag}, CRAG~\cite{yan2024crag}, and analyses of conflicting evidence~\cite{wang2025ragconflicting} likewise arbitrate unreliable retrieval. RAGForensics~\cite{zhang2025traceback} traces poisoned texts after an attack. CoordGuard is complementary. It does not determine which claim is factually correct, but audits whether the retrieved evidence exhibits the structural concentration associated with a coordinated campaign.

\paragraph{Structure-based filtering.}
Work on parametric-memory limits and adaptive retrieval asks whether a model already knows an answer or requires retrieval~\cite{kandpal2023mem,mallen2023strategic}. Such signals are least informative when retrieval is needed precisely because the model lacks the answer. Conventional outlier detection~\cite{aggarwal2017outlier}, relevance pruning, and near-duplicate suppression look for isolated or redundant passages. Coordinated poison can evade these intuitions by using lexically distinct passages that remain semantically aligned and query-relevant. MMR~\cite{carbonell1998mmr} and related diversity objectives reduce redundancy but do not necessarily remove a mutually supportive semantic coalition.

Closest to our method are structure-based retrieval defenses. TrustRAG~\cite{trustrag2025} combines clustering with structural and parametric signals; SeCon-RAG~\cite{seconrag2025} filters and resolves semantic conflicts; and GRADA~\cite{grada2025} reranks an adversarial-document graph. PRA-RAG~\cite{prarag2026} samples passage subsets and uses embedding geometry to derive a robust aggregate under its assumptions, while BiRD~\cite{bird2026} uses bidirectional ranking consistency. On the corpus side, CleanBase~\cite{cleanbase2026} scans the knowledge database offline for semantically tight malicious document groups and GHP~\cite{ghp2026} hardens graph-based reranking for poison with strong query support; at query time, RAGuard~\cite{raguard2025} expands retrieval and filters chunks by reader perplexity, and RevPRAG~\cite{revprag2025} detects attacks from reader activations. CoordGuard instead audits whether the realized top-$k$ contains a query-anchored coordination component, exposes both unconditional and benign-calibrated actions, and explicitly measures their clean-utility cost; it needs no corpus access or poison examples (unlike corpus-side scans) and no reader logits or internals (unlike perplexity- and activation-based filters), so it also covers pool-injected campaigns whose passages are never corpus-indexed.

Polymorphic-sybil poisoning~\cite{polysybil2026} further shows that lexical diversity can preserve coordinated semantic influence and defeat cosine near-duplicate filters. We do not claim cross-protocol dominance. Where a controlled reproduction is feasible (GRADA, SeCon-RAG, BiRD, and TrustRAG re-run under our fixed coordinated-poisoning protocol, Section~\ref{sec:experiments}), we report the comparison directly; for the remaining systems, differing attack construction, retriever interfaces, and action spaces preclude it. Our contribution is the retrieval-concentration analysis and query-anchored component audit under a fixed coordinated-poisoning protocol.

\section{Method: CoordGuard}
\label{sec:method}

\begin{figure*}[t]
\centering
\includegraphics[width=0.75\linewidth]{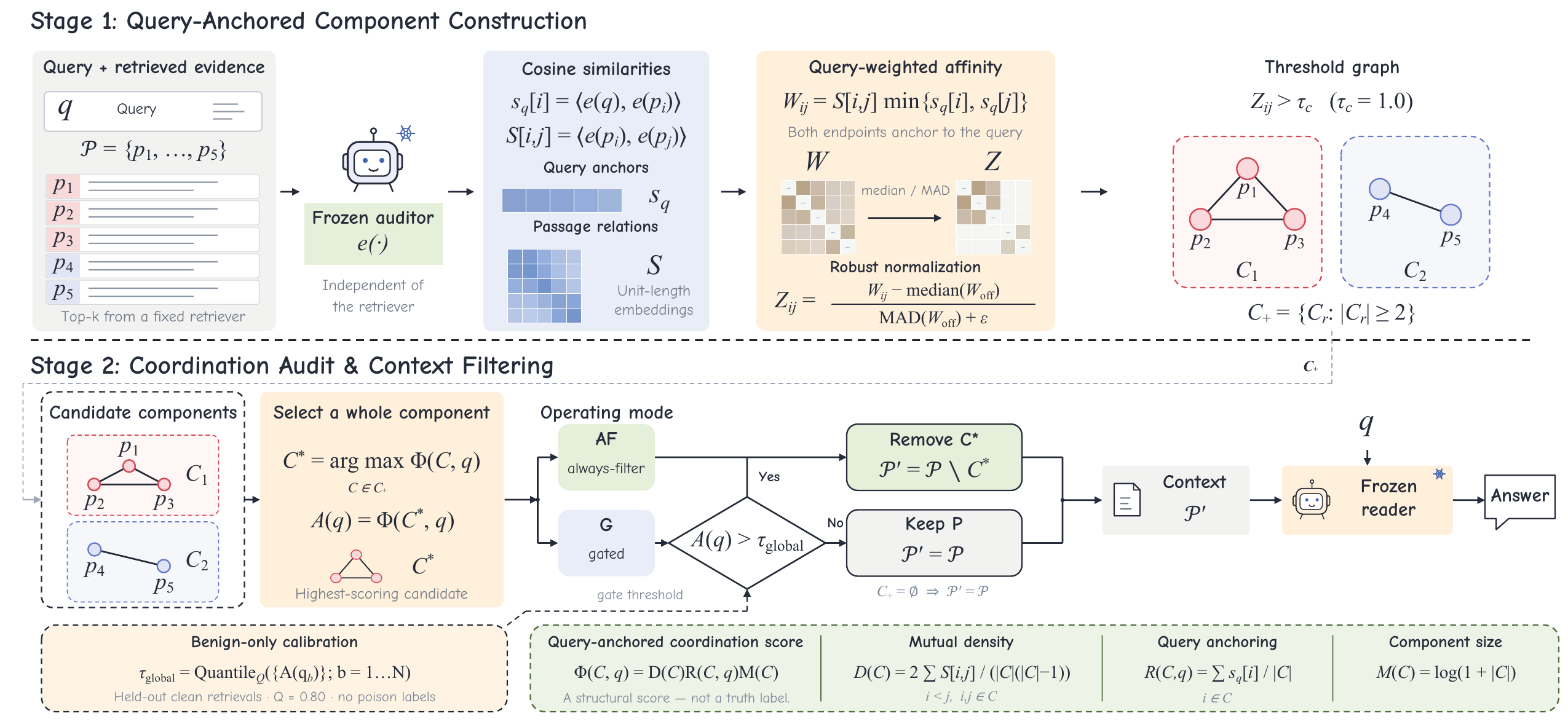}
\caption{CoordGuard overview. Stage 1: query-weighted affinities, robustly normalized and thresholded into candidate components. Stage 2: the top-scoring component supplies $A(q)$; always-filter removes it whenever a component exists, the calibrated gate only at $A(q)>\tau_{\mathrm{global}}$; the filtered context reaches the frozen reader.}
\Description{Two-stage pipeline diagram. Top row: the query and five retrieved passages are embedded by a frozen auditor independent of the retriever; cosine similarities yield the query-weighted affinity matrix W, robustly normalized to Z, and thresholding forms a passage graph whose connected components C1 and C2 make up the candidate set. Bottom row: the components are scored and the highest-scoring C star is selected with audit score A of q; a benign-only calibration sets the gate threshold; always-filter removes C star whenever a component exists while the gate removes it only at unusually high scores; the resulting context reaches the frozen reader, which answers the query.}
\label{fig:framework}
\end{figure*}

CoordGuard audits the structure that an effective coordinated campaign tends to create, namely a multi-passage group that is self-similar, query-relevant, and large enough to influence a small retrieved context. Figure~\ref{fig:framework} summarizes the audit pipeline and its two operating modes; the remainder of this section defines the threat model, the query-anchored component construction and its score, and the two calibration modes.

\subsection{Retrieval concentration and threat model}
\label{sec:threatmodel}

A query $q$ retrieves passages $\mathcal{P}=\{p_1,\ldots,p_k\}$ from a corpus $\mathcal{C}$, and a frozen reader $\pi$ generates an answer $a\sim\pi(\cdot\mid q,\mathcal{P})$. Let $\mathcal{R}_k(q;\mathcal{C})$ denote the top-$k$ retrieval result. After the adversary injects a bounded set $\mathcal{P}_{\mathrm{adv}}$, we define the retrieved poison occupancy
\begin{equation}
 m_k(q)=\left|\mathcal{R}_k(q;\mathcal{C}\cup\mathcal{P}_{\mathrm{adv}})\cap\mathcal{P}_{\mathrm{adv}}\right|,
 \qquad
 \rho_k(q)=\frac{m_k(q)}{k}.
 \label{eq:occupancy}
\end{equation}
A query is \emph{exposed} when $m_k(q)\geq1$ and has a \emph{retrieved adversarial majority} when $m_k(q)>\lfloor k/2\rfloor$. This distinction separates corpus-level sparsity from reader-side dominance. Our headline configuration uses $k{=}5$ and three injected passages; when all three are retrieved, they form a local majority.

Let $e(\cdot)\in\mathbb{R}^d$ be an $\ell_2$-normalized auditor embedding, which may differ from the retrieval encoder. In our primary implementation, passage representations are mean-pooled e5-large-v2 embeddings. We write
\begin{equation}
 s_q[i]=\langle e(q),e(p_i)\rangle,
 \qquad
 S[i,j]=\langle e(p_i),e(p_j)\rangle
 \label{eq:similarities}
\end{equation}
for query--passage relevance and passage--passage similarity.

The adversary selects a target claim $a^*$ and injects passages intended to make the reader emit $a^*$. For evaluation, $a^*$ is constructed to differ from the benchmark answer, but neither CoordGuard nor its calibration procedure uses that answer. The injected passages may vary in wording and semantic coordination while consistently supporting $a^*$. The adversary controls the injected text but not the reader parameters, the fixed retriever, or the auditor. Section~\ref{sec:ablations} evaluates lexical diversity, lower coordination, retrieval optimization, and black-box candidate selection; direct end-to-end optimization against the complete defended pipeline remains outside the present scope.

Effective coordination creates two empirical tendencies. \emph{Retrievability} requires each injected passage to score highly enough against the query to enter the top-$k$. \emph{Mutual reinforcement} requires the retrieved poison passages to remain sufficiently aligned in meaning to advocate the same target. These tendencies imply neither perfect clustering nor theoretical separation. Benign passages can also be mutually similar and query-relevant when they describe the same entity, event, or fact; we call this \emph{benign co-relevance}. Distinguishing adversarial coordination from benign co-relevance is the central difficulty of a per-query gate.

The defender operates at inference time. It observes $q$ and the retrieved passages, can compute an auditor embedding, and may suppress passages before they reach the reader. A held-out collection of benign retrievals is available for calibrating a score threshold, but no poison examples, target answers, correct answers, or reader gradients are used for calibration.

For analysis, we summarize poison--benign separability for auditor $e$ as
\begin{equation}
\mathrm{AUC}_e=\mathrm{AUC}\!\left(
A(q)\mid q{\sim}\mathrm{poisoned}
\;\text{vs.}\;
q{\sim}\mathrm{benign}
\right).
\label{eq:sep}
\end{equation}
This is an evaluation statistic, not a training objective. CoordGuard does not optimize $\mathrm{AUC}_e$ or use poison labels to choose its deployment threshold. Across seven tested embeddings, $\mathrm{AUC}_e$ ranges from $0.657$ to $0.740$ (Appendix~\ref{sec:appendix-ceiling}), which we interpret as protocol-specific empirical separability, not a universal limit.

\subsection{Query-anchored coordination audit}
\label{sec:method-instantiation}

\paragraph{Query-weighted affinity.}
Raw passage similarity alone cannot distinguish a coordinated attack from a dense group of passages about an unrelated topic. CoordGuard therefore weights each pair by the weaker query anchor:
\begin{equation}
W[i,j]=S[i,j]\min\{s_q[i],s_q[j]\}.
\label{eq:affinity}
\end{equation}
The minimum requires both endpoints of an edge to be query-relevant; one strongly anchored passage cannot compensate for a weakly anchored partner. The construction remains reader-independent because it uses only evidence geometry.

The scale and dispersion of $W$ vary across queries. We robustly normalize its off-diagonal entries using the median and median absolute deviation (MAD),
\begin{equation}
Z[i,j]=
\frac{W[i,j]-\operatorname{median}(W_{\mathrm{off}})}
{\operatorname{MAD}(W_{\mathrm{off}})+\epsilon}.
\label{eq:normalize}
\end{equation}
Pairs with $Z[i,j]>\tau_c$ define an undirected graph over the retrieved passages. Its connected components with at least two passages form the candidate set $\mathcal{C}^+$. If $\mathcal{C}^+$ is empty, CoordGuard returns the original context. This prevents a singleton from being removed merely because no multi-passage coordination component exists.

\paragraph{Coordination score.}
Each candidate component is scored by
\begin{equation}
\Phi(C,q)=
\underbrace{\frac{2}{|C|(|C|-1)}
\!\sum_{\substack{i<j\\i,j\in C}}S[i,j]}_{D(C)}
\cdot
\underbrace{\frac{1}{|C|}\sum_{i\in C}s_q[i]}_{R(C,q)}
\cdot
\underbrace{\log(1+|C|)}_{M(C)}.
\label{eq:phi}
\end{equation}
The selected component and audit score are
\begin{equation}
C^*=\arg\max_{C\in\mathcal{C}^+}\Phi(C,q),
\qquad
A(q)=\Phi(C^*,q).
\label{eq:select}
\end{equation}
The product is a scoring heuristic, not a certified likelihood. $D(C)$ measures mutual alignment. $R(C,q)$ discounts components weakly anchored to the query, while $M(C)$ rewards groups large enough to influence a small context. We keep $\Phi$ as the default because it also supplies a continuous audit score for benign-only calibration; it is a calibration-compatible default, not a claim of unique optimality. Under matched removal, largest-, density-, and anchor-based component selection all reach $\mathrm{ASR}_{\mathrm{exp}}=0.13$--$0.18$, while random and lowest-relevance removal remain at $0.48$ and $0.57$ (Table~\ref{tab:countmatch}, Appendix~\ref{sec:appendix-ceiling}).

Selecting $C^*$ does not assert that the component is poisoned. A benign co-relevant group can maximize $\Phi$, which explains both the clean-accuracy cost of unconditional suppression and the imperfect AUC of a per-query gate. Algorithm~\ref{alg:coordguard} gives the inference procedure.

\begin{algorithm}[t]
\small
\caption{CoordGuard audit}
\label{alg:coordguard}
\begin{algorithmic}[1]
\raggedright
\REQUIRE query $q$, passages $\mathcal{P}$, embedding $e$, component threshold $\tau_c$, gate threshold $\tau_{\mathrm{global}}$
\STATE Compute $s_q[i]$ and $S[i,j]$
\STATE $W[i,j]\leftarrow S[i,j]\min\{s_q[i],s_q[j]\}$
\STATE Normalize $W_{\mathrm{off}}$ with its median and MAD to obtain $Z$
\STATE Link $i,j$ when $Z[i,j]>\tau_c$; retain $\mathcal{C}^+=\{C_r:|C_r|\geq2\}$
\IF{$\mathcal{C}^+=\varnothing$}
  \STATE \textbf{return} $\mathcal{P}$
\ENDIF
\STATE $C^*\leftarrow\arg\max_{C\in\mathcal{C}^+}\Phi(C,q)$; $A(q)\leftarrow\Phi(C^*,q)$
\IF{$A(q)>\tau_{\mathrm{global}}$}
  \STATE \textbf{return} $\mathcal{P}\setminus C^*$
\ELSE
  \STATE \textbf{return} $\mathcal{P}$
\ENDIF
\end{algorithmic}
\end{algorithm}

\subsection{Calibration and operating modes}

\paragraph{Benign-only calibration.}
For the gated variant, we compute $A(q)$ on held-out benign retrievals and set $\tau_{\mathrm{global}}$ to their $Q$-th quantile. This controls how often benign queries trigger intervention without requiring poison examples. Increasing $Q$ raises the threshold, reduces intervention, and moves the system toward clean utility at the cost of higher attack success. Calibration is tied to an auditor and retrieval configuration; we do not assume that a numerical threshold transfers unchanged across embedding families.

\paragraph{Primary modes.}
\textsc{CoordGuard-AF} sets $\tau_{\mathrm{global}}=-\infty$ and removes $C^*$ whenever a multi-passage candidate exists. It avoids a per-query poisoned-versus-benign classification decision and provides a robustness-oriented endpoint, but it can remove benign co-relevant components. \textsc{CoordGuard-G} applies the benign-calibrated threshold and intervenes only on queries with unusually high scores. It recovers more clean accuracy but admits more attacks. Neither mode dominates at every deployment cost; together they expose the operating frontier.

Appendix~\ref{sec:appendix-caprefill} tests medoid capping and deeper-rank backfilling. Both recover clean coverage by retaining or reintroducing evidence from the poison coalition, so they are not primary modes.

\paragraph{Complexity and deployment cost.}
Given embeddings, forming the passage similarity matrix costs $O(k^2d)$ and graph construction plus scoring costs $O(k^2)$. On one CPU core with random $d{=}768$ vectors, geometry-only post-processing takes $0.067$ ms median per query at $k{=}5$ and $0.082$ ms at $k{=}10$ (p95 $0.093$ ms). These measurements exclude auditor-embedding inference. This $0.067$\,ms figure is a synthetic algorithmic lower bound (random vectors, single core); on real retrieved passages the deployed geometry cost is $\approx1.3$\,ms/query median, and the full audit (auditor encoding of the retrieved top-$k$ plus geometry) is $\approx33$\,ms/query median and $<70$\,ms at the 95th percentile (Section~\ref{sec:experiments}). Passage embeddings can be precomputed when the corpus and auditor are fixed, but deployment still requires a query embedding and a second passage representation when the auditor differs from the retriever; the $0.067$\,ms value is clustering-and-scoring overhead only, not end-to-end latency.

\section{Experiments}
\label{sec:experiments}

We organize the evaluation around four questions: (RQ1) does query-anchored component filtering reduce exposure-conditional attack success below every evaluated baseline, and at what clean-utility cost? (RQ2) does the geometry signal survive swapping the retrieval and audit spaces, and does it remove more poison than generic outlier filtering at matched intervention? (RQ3) does the reduction transfer across readers, datasets, and attack budgets, or is it a single-configuration artifact? (RQ4) which components carry the gain under count-matched removal, adaptive attacks, and premise-splitting stress?

\subsection{Experimental setup}
\label{sec:setup}

\paragraph{Data, models, and attack.}
We sample $1{,}000$ queries each from Natural Questions (NQ)~\cite{kwiatkowski2019nq} and HotpotQA~\cite{yang2018hotpotqa}, and construct an MS MARCO wrong-claim steering task. Frozen readers are Qwen2.5-7B-Instruct~\cite{qwen25} (primary) and Llama-3.1-8B-Instruct~\cite{llama3}. The anchor retriever is Contriever~\cite{izacard2022contriever}; the primary independent auditor is e5-large-v2~\cite{wang2022e5}. We follow PoisonedRAG's~\cite{poisonedrag2025} targeted knowledge-corruption threat model, while using our own passage-generation and candidate-pool protocol. The three passages are spliced into a clean BM25-top-$50$ pool that Contriever re-ranks, and compete for the top $k{=}5$ under Contriever score instead of being indexed into the corpus. This \emph{pool-injection} model skips BM25's first-stage filter and hence does not disadvantage the adversary, yet leaves deeper BM25 ranks clean by construction; the recovery study uses this property, and we test a corpus-indexed boundary separately. A row is \emph{exposed} when at least one injection enters the top-$k$; the NQ headline pools $3{,}764$ exposed query--seed rows over four retrieval seeds. The parametrically unknown subset contains queries answered incorrectly without retrieval.

\paragraph{Attack-generation protocol.}
For each query, Qwen2.5-7B-Instruct first samples one plausible benchmark-inconsistent target (temperature $0.7$, top-$p$ $0.95$). Conditioned on that fixed target, it then produces three numbered, encyclopedia-style passages in one deterministic decoding call; the passages share the claim but differ in surface form. Reader answers are greedily decoded, so reported variation comes from query sampling and retrieval seeds instead of reader sampling. The paraphrase, low-coordination, retrieval-optimized, gate-aware, and premise-splitting variants use separate prompts or candidate selection and are reported as stress tests, not pooled with the headline protocol. Neither attack generation nor reader inference is used to fit CoordGuard.

\paragraph{Metrics and baselines.}
We report clean and poisoned accuracy and exposure-conditional attack success $\mathrm{ASR}_{\mathrm{exp}}$; unconditional end-to-end ASR additionally counts unexposed targeted rows. Gold- and target-matching both use substring containment after normalization, so a prediction containing both strings counts in both columns and the two columns need not sum to one. Table~\ref{tab:recovery} reports a mutually exclusive decomposition; directly measured both/neither string rates account exactly for the overlap (Appendix~\ref{sec:appendix-metric}). Geometric diagnostics include $\mathrm{AUC}_e$, passage discrimination, and \emph{Det-all}, the fraction of exposed attacks for which all retrieved poison is removed. Baselines are closed-book and vanilla RAG; RobustRAG~\cite{roburstrag2024}; ReliabilityRAG~\cite{reliabilityrag2025}; MMR~\cite{carbonell1998mmr} and cosine deduplication; CAD~\cite{shi2024cad} and CoCoA~\cite{wang2025cocoa}; TrustRAG-S1/S2~\cite{trustrag2025}; and relevance and isolation outliers. Each method is run on its own clean-side pipeline. In our short-answer interface, RobustRAG's plurality and ReliabilityRAG's maximum independent set of the binary answer-mismatch graph coincide row-by-row; this protocol equivalence does not imply that the full methods are identical. TrustRAG-S1/S2 use the same four retrieval seeds as CoordGuard and are paired query-by-query; CAD and CoCoA use method-specific decoding interfaces and lack per-query multi-seed records, so those comparisons remain descriptive.

\paragraph{Reproducibility.}
Four-seed CoordGuard and TrustRAG intervals use a two-level bootstrap over query IDs and retrieval-seed rows ($B{=}2000$); single-seed runs use query bootstrap. TrustRAG-S1/S2 are paired query-by-query with CoordGuard and use the same bootstrap; baselines without per-query multi-seed records are reported as point estimates, without bootstrap-based significance claims. We set $\tau_c{=}1.0$, $k{=}5$, and $n_{\mathrm{poison}}{=}3$ using held-out benign retrievals; gate thresholds are selected only from benign calibration data. Experiments use a shared $3{\times}$A800 80GB server, with the geometry score computed on CPU ($\approx1.3$\,ms/query median). The full audit (auditor encoding of the retrieved top-$k$ plus geometry) costs $\approx33$\,ms/query median and $<70$\,ms at the $95$th percentile, under $10\%$ of reader generation latency; it intervenes on $92\%$ of poisoned and $90\%$ of benign queries, removes a mean of $2.4$--$2.5$ of $5$ passages, and blanks the full context in under $0.2\%$ of queries.

\subsection{Robustness, utility, and retrieval concentration (RQ1)}
\label{sec:main}

Table~\ref{tab:main} gives the four-seed NQ comparison. Vanilla RAG reaches $\mathrm{ASR}_{\mathrm{exp}}=0.576$. RobustRAG and ReliabilityRAG reduce this to $0.512$ but also reduce clean accuracy to $0.216$ in the evaluated shared protocol. Generic diversity filters and context-amplifying decoding barely move the attack. TrustRAG-S1 reaches clean $0.299$ at $\mathrm{ASR}_{\mathrm{exp}}=0.313$ $[0.285,0.339]$ and TrustRAG-S2 reaches clean $0.217$ at $\mathrm{ASR}_{\mathrm{exp}}=0.270$ $[0.244,0.297]$, each on the same four seeds. Paired query-by-query against CoordGuard-AF, the attack-success gap is $\Delta\mathrm{ASR}=-0.145$ $[-0.171,-0.119]$ versus TrustRAG-S1 and $-0.100$ $[-0.125,-0.073]$ versus TrustRAG-S2, with $95\%$ CIs excluding zero. TrustRAG-S2's internal-knowledge arbitration carries a clean cost ($0.217$, versus $0.299$ for S1) without recovering robustness, so it is dominated on both axes by the calibrated Contriever gate. The closest concurrent retrieval-time defense, BiRD~\cite{bird2026}, is competitive at the high-robustness end. Reproduced under the same protocol, with the retriever's Contriever embeddings and a benign-calibrated threshold, it reaches $\mathrm{ASR}_{\mathrm{exp}}=0.182$ at clean $0.222$, against $0.167$ at $0.192$ for CoordGuard-AF. CoordGuard's advantage lies at high utility. At matched clean accuracy of roughly $0.30$, BiRD reaches only $0.479$ while the calibrated gate stays between $0.27$ and $0.30$.

With e5 geometry, \textsc{CoordGuard-AF} reaches $\mathrm{ASR}_{\mathrm{exp}}=0.167$ (95\% CI $[0.144,0.190]$) and $0.138$ on parametrically unknown queries. Poisoned accuracy rises from $0.158$ to $0.200$, so the reduction is not merely a shift to a third wrong answer. Clean accuracy decreases from $0.328$ to $0.192$. Contriever geometry gives a less aggressive endpoint (clean $0.254$/$\mathrm{ASR}_{\mathrm{exp}}=0.211$), and its benign-calibrated $Q{=}0.80$ gate reaches $0.315/0.331$ with no significant clean-accuracy cost (paired $\Delta{=}{-}1.2$pp, $95\%$ CI $[-2.8,+0.4]$, crossing zero).

\paragraph{Closest structure-based defenses under one protocol.}
We also reproduce the two closest published structure-based filters on the same single-seed dump (977 exposed NQ rows, Qwen2.5-7B reader, e5 auditor, paired bootstrap $B{=}2000$). GRADA~\cite{grada2025} reranks an inter-document similarity graph by PageRank, interpolating graph score with query relevance through a weight $\alpha$; SeCon-RAG~\cite{seconrag2025} splits the context into two clusters and keeps the more self-consistent one. CoordGuard filtering reaches $\mathrm{ASR}_{\mathrm{exp}}=0.153$ $[0.130,0.175]$, against $0.245$ $[0.218,0.271]$ for GRADA at $\alpha{=}0$ and $0.352$ $[0.322,0.382]$ at $\alpha{=}1$, and $0.407$ $[0.377,0.438]$ for SeCon-RAG; no interval overlaps CoordGuard's. The failure modes are diagnostic. GRADA degrades monotonically as its query-relevance weight grows, because coordinated poison is query-relevant by construction, the exact property CoordGuard exploits. SeCon-RAG retains the most coherent cluster, which under this attack is the coalition itself.

On HotpotQA, e5 AF reduces $\mathrm{ASR}_{\mathrm{exp}}$ from $0.735$ to $0.309$ versus $0.429$ for TrustRAG-S1, while incurring only a $1.6$-point clean loss (Table~\ref{tab:mainhp}, Appendix~\ref{sec:appendix-ceiling}). This contrast with NQ shows that utility cost depends on benign evidence structure instead of filtering aggressiveness alone. Consistent with this, on benign NQ $80\%$ of the queries where CoordGuard lowers accuracy are single-source-evidence queries whose sole supporting passages form the detected cluster; near-duplicates account for only $1\%$, so the cost is not recoverable by deduplication. It is the flip side of cluster-coherence detection, concentrated where the answer lives in one tightly-clustered evidence source.

\paragraph{Corpus-indexed end-to-end.}
With the attack passages \emph{indexed into the BM25 corpus}, exposure falls from $0.941$ to $0.766$ and the adversarial-majority rate from $0.529$ to $0.199$, yet vanilla $\mathrm{ASR}_{\mathrm{exp}}$ stays at $0.513$. Retrieval shields some rows entirely, and the rows it does not shield are compromised as before. CoordGuard-AF cuts attack success to $0.178$ ($-33.5$ points). The benign-calibrated gate is the better mode here. At $Q{=}0.80$ it reaches $0.333$ at clean $0.291$, dominating RobustRAG and ReliabilityRAG ($0.417$ at clean $0.216$) on both axes in an independently regenerated four-seed run (Table~\ref{tab:ci}, Appendix~\ref{sec:appendix-ceiling}). The signal does not weaken under corpus indexing. Retrieval already protects some rows, which is precisely the trade the gate makes.

\begin{table}[t]
\centering
\scriptsize
\setlength{\tabcolsep}{2.35pt}
\renewcommand{\arraystretch}{0.82}
\caption{NQ results (four retrieval seeds, $3{,}764$ exposed rows; TrustRAG rows use the same seeds, paired query-by-query with CoordGuard). Exposure $0.941$; adversarial-majority rate $0.529$; unconditional end-to-end ASR $0.547$ (vanilla) and $0.162$ (e5 AF). $^\dagger$Method-specific decoding; CAD/CoCoA comparisons are descriptive.}
\label{tab:main}
\begin{tabular}{lccc}
\toprule
Method & Clean & Pois. Acc & $\mathrm{ASR}_{\mathrm{exp}}$ \\
\midrule
Closed-book (same rows) & 0.155 & 0.155 & 0.138 \\
Vanilla RAG & \textbf{0.328} & 0.158 & 0.576 \\
RobustRAG & 0.216 & 0.149 & 0.512 \\
ReliabilityRAG & 0.216 & 0.149 & 0.512 \\
MMR & 0.307 & 0.171 & 0.543 \\
Cosine deduplication & 0.328 & 0.170 & 0.560 \\
CAD$^\dagger$ & 0.303 & 0.169 & 0.601 \\
CoCoA$^\dagger$ & 0.302 & 0.161 & 0.555 \\
TrustRAG-S1 & 0.299 & 0.187 & 0.313\,[0.285,0.339] \\
TrustRAG-S2 & 0.217 & 0.167 & 0.270\,[0.244,0.297] \\
\midrule
CoordGuard-AF (Contriever) & 0.254 & \textbf{0.224} & 0.211 \\
CoordGuard-AF (e5) & 0.192 & 0.200 & \textbf{0.167}\,[0.144,0.190] \\
CoordGuard-G ($Q{=}0.80$) & 0.315 & 0.212 & 0.331 \\
\bottomrule
\end{tabular}
\end{table}

\paragraph{Poison multiplicity and consistency breakdown.}
The mechanism becomes clearer after stratifying by realized poison occupancy $m_k$ (Figure~\ref{fig:heatmap}b). RobustRAG and ReliabilityRAG are effective with one retrieved poison passage but rise from $\mathrm{ASR}_{\mathrm{exp}}=0.228$ at $m_k{=}1$ to $0.519$ at $m_k{=}2$ and $0.591$ at $m_k{=}3$. The degradation at $m_k{=}2$ does not contradict their certificates. Sufficient conditions depend on the number and consistency of useful benign responses, not passage count alone. Empirically, two mutually supportive poison passages can become the largest coherent group when the remaining benign evidence is noisy or inconsistent. CoordGuard has the complementary profile. A lone poison forms no multi-passage component, whereas $\mathrm{ASR}_{\mathrm{exp}}$ falls to $0.124/0.141$ once coordination is present (Figure~\ref{fig:heatmap}b, Appendix~\ref{sec:appendix-ceiling}). This observational stratification lets $m_k$ confound with query difficulty; we isolate the effect by holding the query and clean evidence fixed and varying only $m_k\in\{0,1,2,3\}$ (four seeds, $\approx510$--$530$ queries/seed). Attack strength rises monotonically ($\mathrm{ASR}_{\mathrm{exp}}$ $0.064{\to}0.484{\to}0.543{\to}0.578$), while CoordGuard-AF shows the inverted profile, weakest at $m_k{=}1$ ($0.213$ $[0.171,0.254]$, where a lone poison joins no component) but $0.067$ and $0.081$ at $m_k{=}2$ and $3$. Component formation, not retrieval exposure, drives the breakdown (Figure~\ref{fig:robustness}b).

\subsection{Geometry and transfer (RQ2--RQ3)}
\label{sec:mechanism}

\begin{figure*}[t]
\centering
\includegraphics[width=0.75\textwidth]{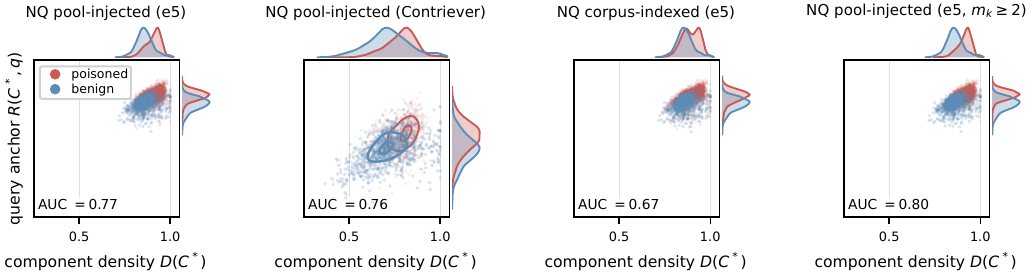}
\caption{Coordination footprint in the $(D,R)$ plane (component-forming rows). Red: poisoned; blue: poison-free retrieval of the same NQ queries and auditor; lines: $50\%$/$90\%$ KDE contours; top/right: marginal KDEs; AUC: query-level $A(q)$ separation. (a) e5, four seeds ($3{,}475$ rows); (b) Contriever, one seed; (c) corpus-indexed, four seeds; (d) e5, $m_k{\geq}2$.}
\Description{Four joint scatter panels, each with marginal density strips on the top and right. In each, red points (poisoned retrieval) concentrate toward higher similarity and query affinity than blue points (benign retrieval), with partial overlap; annotated AUC values range from 0.67 to 0.80.}
\label{fig:geometry}
\end{figure*}

The reader retriever and auditor are decoupled. Replacing Contriever audit geometry with e5 improves AF $\mathrm{ASR}_{\mathrm{exp}}$ from $0.211$ to $0.167$ and passage-level discrimination from $0.671$ to $0.774$, while the seven-embedding sweep is non-monotonic ($\mathrm{AUC}_e=0.657$--$0.740$; Appendix~\ref{sec:appendix-ceiling}). Figure~\ref{fig:geometry} shows the footprint directly. Components selected from poisoned retrieval concentrate in the dense, query-anchored corner of the $(D,R)$ plane under both auditors (query-level AUC $0.76$--$0.77$), weaken but persist under corpus indexing ($0.67$), and strengthen with realized coordination ($0.80$ at $m_k{\geq}2$), while genuinely overlapping benign co-relevance. Crossed retrieval--auditor experiments reduce $\mathrm{ASR}_{\mathrm{exp}}$ in all four cells (Contriever retrieval with a Contriever/e5 auditor attains $0.201/0.179$; e5 retrieval $0.140/0.128$). Strengthening the first-stage retriever itself, the \texttt{ms-marco-MiniLM-L-6-v2} cross-encoder over the same BM25 top-50, makes the coordinated attack \emph{harder}. Vanilla $\mathrm{ASR}_{\mathrm{exp}}$ falls $0.558{\to}0.470$ $[0.451,0.490]$ (four seeds), yet CoordGuard-AF holds at $0.139$ $[0.122,0.156]$; whichever retriever surfaces the coalition, the coalition must still form a dense query-anchored component, so the defense transfers to a stronger, more realistic retriever. Relevance and isolation outliers remove all poison in only $8\%/17\%$ of attacks versus $73\%$ for component filtering, supporting the dense-inlier diagnosis.

\begin{figure}[b]
\centering
\includegraphics[width=0.8\linewidth]{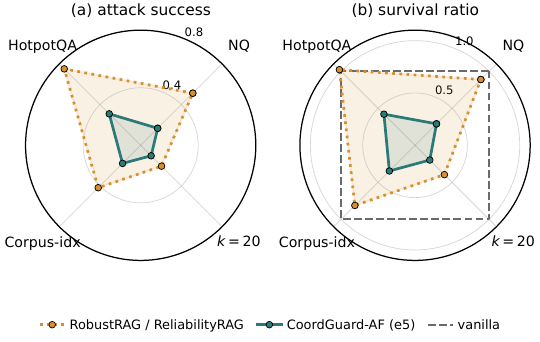}
\caption{Radar of attack success for the three methods with complete coverage of the four settings. (a) $\mathrm{ASR}_{\mathrm{exp}}$ (toward the center is better). (b) Survival ratio vs vanilla (dashed ring at $1$; above $1$ is worse than undefended, as the aggregators reach $1.02$ on HotpotQA). RobustRAG and ReliabilityRAG coincide (one series); Corpus-idx and $k{=}20$ independently regenerated.}
\Description{Two radar charts with four axes labeled NQ, HotpotQA, Corpus-idx, and k=20. Panel a shows RobustRAG equal to ReliabilityRAG outermost and CoordGuard-AF closest to the center. Panel b shows the survival-ratio view with a dashed vanilla ring at 1; the aggregation baselines sit at 0.81 to 1.02 while CoordGuard-AF sits at 0.20 to 0.42.}
\label{fig:radar}
\end{figure}

The reduction transfers across readers and tasks. e5 AF changes $\mathrm{ASR}_{\mathrm{exp}}$ from $0.576\to0.167$ on Qwen-NQ, $0.528\to0.145$ on Llama-NQ, $0.735\to0.309$ on HotpotQA, and $0.363\to0.158$ on MS MARCO steering. A newer reader family replicates the ranking. On Qwen3.5-9B~\cite{qwen35} (single seed, 944 exposed rows; greedy short-answer decoding with thinking mode disabled), e5 always-filtering holds $\mathrm{ASR}_{\mathrm{exp}}=0.145$ and the Contriever auditor $0.190$, while RobustRAG and ReliabilityRAG sit at $0.477$ and vanilla at $0.528$, each within $1$--$3$ points of its Qwen2.5 counterpart, so the geometric signal does not erode on a newer reader. Clean cost ranges from about $1$ to $14$ points across datasets; what transfers is the robustness effect, with the operating point set per deployment by the benign-calibrated gate. Budget and $k$ sweeps, including $k{=}20$, where AF reaches $0.103$ with near-maximal removal, are in Appendix~\ref{sec:appendix-scale} (Figure~\ref{fig:robustness}d). Figure~\ref{fig:radar} replays the attack-success landscape as a radar over the four settings with complete method coverage.

\subsection{Matched controls, calibration, and adaptation (RQ4)}
\label{sec:ablations}

\paragraph{What drives the gain?}
Count-matched deletion separates selection quality from intervention volume (Table~\ref{tab:countmatch}, Appendix~\ref{sec:appendix-ceiling}). On the matched subset, every component-targeted selector reaches $\mathrm{ASR}_{\mathrm{exp}}=0.13$--$0.18$ and removes all poison in $0.75$--$0.85$ of attacks, whereas random and lowest-relevance deletion stay at $0.47/0.57$ ($\text{Det-all}\le0.21$). No single selector dominates $\Phi$. The anchor-only rule edges it by $1.0$pp ($\Delta\mathrm{ASR}_{\mathrm{exp}}=-0.010$ $[-0.018,-0.003]$) at slightly lower clean accuracy ($0.179$ vs $0.192$); the largest-component rule ties ($+0.006$, n.s.); density-only and similarity-contribution fall behind ($+0.017$/$+0.041$). The active unit is the query-weighted component, not the $\Phi$ score itself. $\Phi$ is retained as the default; it is the continuous score under which the benign-calibrated gate stays robust to score-aware adaptation (Table~\ref{tab:adaptive}, Appendix~\ref{sec:appendix-ceiling}), and removing its query-relevance term raises $\mathrm{ASR}_{\mathrm{exp}}$ from $0.167$ to $0.188$.

\begin{figure}[ht]
\centering
\includegraphics[width=0.75\linewidth]{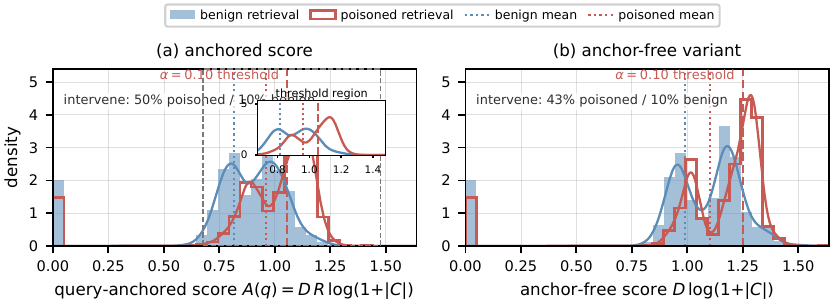}
\caption{Audit-score distributions and the benign-calibrated threshold (regenerated four-seed NQ dumps; benign $n{=}1{,}000$ filled, attack $n{=}3{,}751$ outline; dotted: class means; dashed: benign $90$th percentile; inset in (a) zooms the threshold). (a) Query-anchored $A(q)$; (b) anchor-free variant.}
\Description{Two histograms with dotted class-mean lines; the poisoned distribution sits further right of the benign threshold in the anchored panel than in the anchor-free panel, with intervention rates annotated, and panel (a) carries a small inset zoom on the threshold region.}
\label{fig:gatehist}
\end{figure}

\begin{figure*}[ht]
\centering
\includegraphics[width=0.75\textwidth]{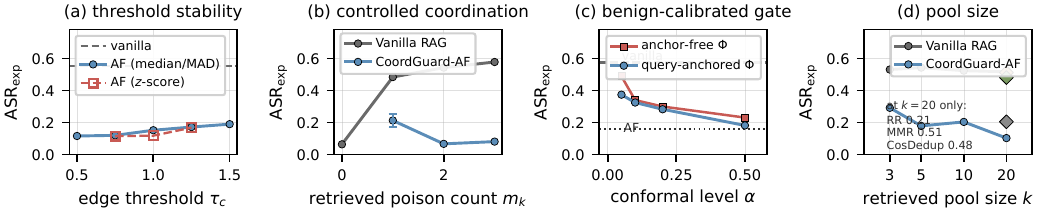}
\caption{Robustness across operating parameters ($\mathrm{ASR}_{\mathrm{exp}}$; lower is better). (a) $\tau_c$ over a $3\times$ range, median/MAD and $z$ normalization: no cliff. (b) Controlled coordination, query and evidence fixed; bar: $95\%$ CI. (c) Conformal frontier, query-anchored vs.\ anchor-free score; dashed/dotted: vanilla/AF references. (d) Pool size $k$ (log scale); diamonds: $k{=}20$-only baselines.}
\Description{Four line plots. Panel a shows attack success nearly flat across the edge threshold for both normalizations. Panel b contrasts a rising vanilla curve with the inverted CoordGuard curve. Panel c shows the anchored score dominating the anchor-free variant at every conformal level. Panel d shows CoordGuard falling further below vanilla as the pool grows.}
\label{fig:robustness}
\end{figure*}

\paragraph{Calibration.}
The Contriever gate traces a monotone frontier from always-filter ($0.254$ clean, $\mathrm{ASR}_{\mathrm{exp}}=0.211$) through the benign-only \emph{clean-budget} point, the most aggressive threshold satisfying a clean-accuracy floor of $0.30$, at $0.302$/$0.267$, to $Q{=}0.80/0.90/0.95$ gates at $0.315$/$0.331$, $0.324$/$0.405$, and $0.330$/$0.444$. The clean-budget point improves on four-seed TrustRAG-S1 on both axes with matched seed coverage; the Contriever auditor and TrustRAG's K-means filtering remain different interfaces, so we read the curve as a family of operating points, not a paired Pareto claim. Recasting the threshold as a conformal $p$-value makes the false-filter guarantee explicit. With $n{=}500$ held-out benign calibration queries, $p(q){=}(1{+}\#\{j{:}A_j{\geq}A(q)\})/(n{+}1)$, and the gate intervenes whenever $p(q){\leq}\alpha$; the clean-query intervention rate tracks $\alpha$ ($0.054/0.096/0.206/0.308$ at $\alpha{=}0.05/0.10/0.20/0.30$; four seeds, e5 score; Figure~\ref{fig:gatehist}). The query-anchored score dominates its anchor-free variant at every $\alpha$, by up to $11.5$ points where the gate is most conservative ($\mathrm{ASR}_{\mathrm{exp}}$ $0.374$ vs.\ $0.489$ at $\alpha{=}0.05$). At $\alpha{=}0.10$ the conformal gate operates at $\mathrm{ASR}_{\mathrm{exp}}{=}0.325$ and clean accuracy $0.308$, intervening on $51\%$ of attacked and $10\%$ of clean queries, the CB point, with a distribution-free threshold. The construction is also insensitive to its remaining hyperparameters. AF $\mathrm{ASR}_{\mathrm{exp}}$ stays within $0.116$--$0.190$ across a $3\times$ range of $\tau_c$ and both MAD and $z$ normalizations, with clean accuracy in a $1.5$-point band and no cliff (Figure~\ref{fig:robustness}a; Appendix~\ref{sec:appendix-scale}). Poison similarity edges sit far above $\tau_c$, so component formation is threshold-stable.

\paragraph{Adaptive stress tests and recovery limits.}
Paraphrasing increases detectability, whereas a low-coordination generator lowers AUC to $0.690$ and raises defended $\mathrm{ASR}_{\mathrm{exp}}$ to $0.290$. In the strongest end-to-end black-box test, candidate selection lowers $A(q)$ from $0.86$ to $0.81$ on $937$ exposed queries; AF $\mathrm{ASR}_{\mathrm{exp}}$ rises from $0.180$ to $0.219$ but remains well below the undefended $0.590/0.588$. On the same adapted set the calibrated gate ($Q{=}0.80$) intervenes on roughly half of exposed queries, so its defended $\mathrm{ASR}_{\mathrm{exp}}$ ($0.348$) is about twice AF; the score-aware attack leaves the intervention rate essentially unchanged ($\approx0.46$) and does not raise gate $\mathrm{ASR}_{\mathrm{exp}}$ ($0.348\to0.303$, bootstrap CIs overlap). It only breaches the gate at aggressive thresholds, where score-lowering dodges intervention ($Q{=}0.95$: $0.33\to0.13$, $\mathrm{ASR}_{\mathrm{exp}}\to0.505$); the $Q{=}0.90$ rise is marginal (CIs overlap). A \emph{multi-objective} variant that places the reader in the candidate-selection loop, keeping hijack success while lowering $A(q)$, strengthens the undefended attack ($0.59\to0.74$, two seeds) yet only raises defended AF $\mathrm{ASR}_{\mathrm{exp}}$ to $0.23$, the same ceiling, and $A(q)$ falls only mildly ($0.87\to0.83$). Medoid capping and deeper-rank backfilling restore clean coverage by preserving or reintroducing poison, raising $\mathrm{ASR}_{\mathrm{exp}}$ to $0.45$--$0.49$ (Appendix~\ref{sec:appendix-caprefill}).

\paragraph{From suppression to recovery.}
Always-filtering suppresses the coalition but does not itself recover the gold answer. Pooling the four seeds, on vanilla-hijacked rows CoordGuard-AF sends $0.210$ to gold, $0.122$ remain on the target, and $0.608$ drift to a third answer (Table~\ref{tab:recovery}, Appendix~\ref{sec:appendix-ceiling}). We test \emph{refilling} the deleted slots from a poison-free source. Under pool injection, the adversary's passages are not corpus-indexed, so deeper BM25 ranks are clean by construction (zero of $943$ attacked rows share a backfill candidate with any poison passage), and refill cannot re-introduce this attack. Refilling by query relevance after full cluster removal raises gold recovery on hijacked rows from $0.210$ to $0.402$ (paired query-bootstrap, $n{=}1{,}945$, $95\%$ CI $[{+}0.169,{+}0.213]$) and gold accuracy on all attacked rows from $0.208$ to $0.364$, while attack success falls slightly ($0.122\to0.083$). The cap-and-refill variant of Appendix~\ref{sec:appendix-caprefill}, in contrast, retains a component medoid and backfills from the retrieved ranking, resurfacing poison ($\mathrm{ASR}_{\mathrm{exp}}\approx0.45$). The repulsive counter-retrieval score (Table~\ref{tab:recovery}) matches random backfill ($0.231$ vs $0.237$ gold). In pool injection, poison and gold evidence share the query-anchor neighborhood, so repelling the deleted component also repels the clean answer. Under corpus-indexed poison a query's own passages resurface in $\sim$75\% of rows but a co-attacker's reaches the backfill range in only $\sim$3--5\%, so relevance refill still dominates the repulsive score (gold $0.403$ vs $0.232$, $95\%$ CI $[{+}0.124,{+}0.218]$) at equal ASR. Only when the deleted coalition is forced into the backfill does relevance refill resurface poison ($0.759$) while the repulsive score avoids it ($0.097$). The repulsive term earns its keep only when the refill source holds coalition-resembling passages. On clean queries where always-filtering deletes a benign component, relevance refill restores gold accuracy to $0.279$, above vanilla ($0.219$) and always-filtering ($0.145$), so recovery lowers the clean-utility cost instead of adding to it.

\paragraph{Premise-splitting stress.}
A natural falsifier inverts only the inter-passage alignment. Under premise-splitting (PS-2), the three injections still assert one benchmark-inconsistent target but via mutually divergent, surface-distinct support. This \emph{strengthens} the attack on the undefended reader (vanilla $\mathrm{ASR}_{\mathrm{exp}}$ $0.558\to0.678$, positive in $4/4$ seeds) yet leaves CoordGuard-AF flat ($0.151\to0.144$; CIs cross zero); the coordination gap widens $0.407\to0.534$. The component does not dissolve. $A(q)$ rises ($0.950\to1.010$), and components of size $\geq2$ form in $0.945$ of rows, because each injection still mentions the query entity, so query-weighted affinity stays dense regardless of whether the support agrees. CoordGuard reads \emph{relevance-anchored geometry, not answer consensus}. Splitting the target itself (PS-1, each injection a different wrong answer) still forms the component ($0.944$ of rows, $A(q){=}1.056$); with no shared target, $\mathrm{ASR}_{\mathrm{exp}}$ is ill-defined and we report only the stored-wrong-answer lower bound ($0.095$). Base coordinated on these seeds falls within the headline CIs; the premise-split claim rests on the split$-$base paired difference, which controls for query sampling (Table~\ref{tab:premise}, Appendix~\ref{sec:appendix-ceiling}).

\FloatBarrier

\section{Conclusion}
\label{sec:conclusion}
\enlargethispage{5\baselineskip}

Globally sparse poisoning can become locally dominant after retrieval. Attack success and consistency-aggregation failure rise with realized poison occupancy, and CoordGuard becomes effective once a multi-passage coalition forms, auditing this footprint by constructing dense, query-anchored evidence components before generation. Across four retrieval seeds, e5 always-filtering reduces exposure-conditional ASR to $0.167$, and the reduction transfers across the tested readers, datasets, and retriever--auditor pairs. When the adversary is corpus-indexed, always-filtering still cuts vanilla ASR by a third, and the benign-calibrated gate dominates aggregation defenses on both axes; refilling deleted context from poison-free deeper ranks then restores gold accuracy on attacked rows ($0.21\to0.40$).

Geometry audits coordination, not truth. Legitimate co-relevant passages can resemble coordinated poison. Deployment requires benign calibration from the intended retrieval distribution, renewed after any change to retriever, auditor, corpus, or top-$k$, with monitoring of intervention rate and removed-component size; retrieved evidence should be treated as an auditable attack surface, combined with provenance and content verification.

\paragraph{Future work.} Three directions follow. First, the geometry-only separability ceiling ($\mathrm{AUC}_e$ $0.657$--$0.740$) invites orthogonal signals (provenance, content verification, or corpus-side usage statistics) that could raise the calibrated gate without retraining the reader. Second, oracle-free clean-cost recovery remains open. The tested knowability signals route only a small fraction of queries better than chance, and a stronger signal would close most of the remaining utility gap. Third, our threat model excludes end-to-end co-optimization against the complete defended pipeline and readers beyond $8$B parameters; a jointly optimized adversary and larger scales would further pin down the audit's boundary.

\FloatBarrier
\clearpage

\bibliographystyle{ACM-Reference-Format}
\setlength{\bibsep}{0pt}
\bibliography{references}

\appendix
\section{Empirical separability across auditor embeddings}
\label{sec:appendix-ceiling}

\begin{figure*}[t]
\centering
\includegraphics[width=0.53\textwidth]{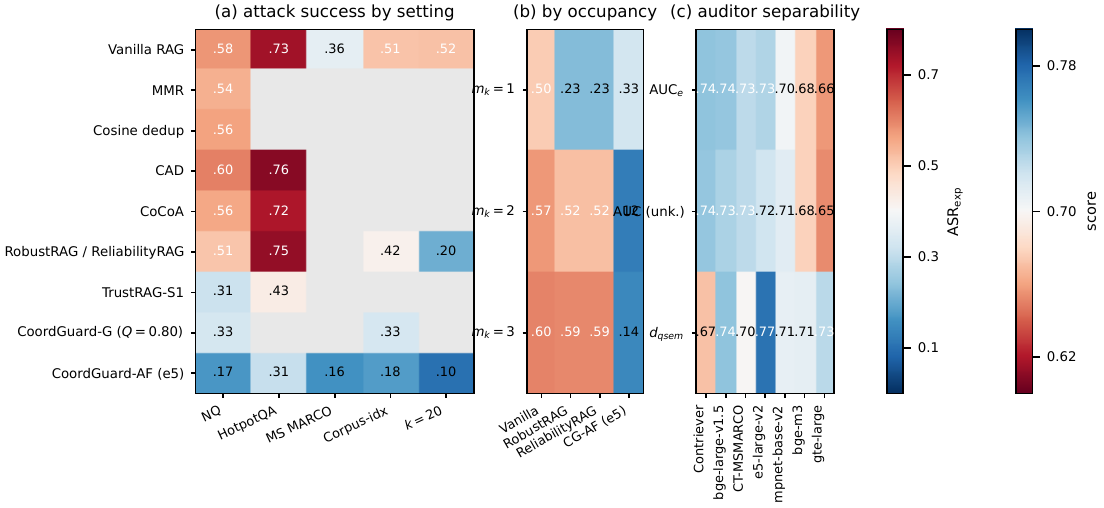}
\caption{Attack-success landscape and auditor separability. (a) $\mathrm{ASR}_{\mathrm{exp}}$ by method and setting (NQ four-seed; others transfer or independently regenerated runs); gray: not evaluated; blue is better. (b) $\mathrm{ASR}_{\mathrm{exp}}$ by realized poison occupancy. (c) Auditor separability across seven embedding spaces.}
\Description{Three heatmaps. Panel a shows attack success falling toward the CoordGuard rows across five setting columns, with gray unevaluated cells. Panel b shows aggregation baselines degrading with poison occupancy while CoordGuard improves. Panel c shows separability scores across seven auditor embeddings.}
\label{fig:heatmap}
\end{figure*}

Across seven auditor embeddings, query-level poison--benign separation spans $\mathrm{AUC}_e=0.657$--$0.740$ (Figure~\ref{fig:heatmap}c). Passage discrimination and query-level separation are not monotone because stronger spaces may tighten both adversarial and benign co-relevant components. A multi-statistic logistic gate adds only $0.003$ AUC without interactions and $0.015$ with interactions. These values are protocol-specific, not a theoretical ceiling. AF does not require per-query classification; it instead pays clean-accuracy cost by intervening on benign queries.

Tables~\ref{tab:mainhp}--\ref{tab:premise} report main-text results (the HotpotQA headline, corpus-indexed end-to-end comparison, count-matched selection, adaptive stress tests, and premise-splitting) placed in this appendix for space. Figure~\ref{fig:heatmap} condenses the attack-success landscape: panel (a) places every evaluated method against the five settings (gray cells: not evaluated under that protocol; CAD and CoCoA use method-specific decoding, so their cells are descriptive), panel (b) the $\mathrm{ASR}_{\mathrm{exp}}$ stratification by realized poison occupancy, and panel (c) auditor separability across the seven embedding spaces.

\begin{table}[tbp]
\centering
\scriptsize
\setlength{\tabcolsep}{2.8pt}
\renewcommand{\arraystretch}{0.88}
\caption{HotpotQA results on exposed rows (one seed); both matchers run over the same rows and are non-exclusive (see setup), so the two columns need not sum to one.}
\label{tab:mainhp}
\begin{tabular}{lccc}
\toprule
Method & Pois. Acc & $\mathrm{ASR}_{\mathrm{exp}}$ & $\mathrm{ASR}_{\mathrm{exp}}$ (unk.) \\
\midrule
Vanilla RAG & 0.281 & 0.735 & 0.756 \\
RobustRAG & 0.286 & 0.751 & 0.762 \\
CAD & 0.286 & 0.764 & 0.794 \\
CoCoA & 0.328 & 0.722 & 0.772 \\
TrustRAG-S1 & 0.290 & 0.429 & 0.404 \\
CoordGuard-AF (e5) & 0.284 & \textbf{0.309} & \textbf{0.242} \\
\bottomrule
\end{tabular}
\end{table}

\begin{table}[tbp]
\centering
\scriptsize
\setlength{\tabcolsep}{2.6pt}
\renewcommand{\arraystretch}{0.88}
\caption{Corpus-indexed end-to-end (NQ--Qwen, e5 auditor, four regenerated retrieval seeds; attack passages in the BM25 corpus; $3{,}017$ exposed rows, exposure $0.766$, majority rate $0.199$). Clean columns from a poison-free dump; $Q$ gates are benign-calibrated quantiles (intervening on $37/32/28\%$ of attacked and $20/15/10\%$ of benign queries at $Q{=}0.80/0.85/0.90$).}
\label{tab:ci}
\begin{tabular}{lccc}
\toprule
Method & Clean & Pois. Acc & $\mathrm{ASR}_{\mathrm{exp}}$ \\
\midrule
Vanilla RAG & \textbf{0.325} & \textbf{0.211} & 0.513 \\
RobustRAG / ReliabilityRAG & 0.216 & 0.175 & 0.417 \\
CoordGuard-AF (e5) & 0.192 & 0.191 & \textbf{0.178} \\
CoordGuard-G ($Q{=}0.80$) & 0.291 & 0.206 & 0.333 \\
CoordGuard-G ($Q{=}0.85$) & 0.298 & 0.211 & 0.349 \\
CoordGuard-G ($Q{=}0.90$) & 0.307 & 0.213 & 0.365 \\
\bottomrule
\end{tabular}
\end{table}

\begin{table}[t]
\centering
\scriptsize
\setlength{\tabcolsep}{3.1pt}
\renewcommand{\arraystretch}{0.88}
\caption{Count-matched selection (NQ-Qwen, e5, four retrieval seeds). Every row removes CoordGuard's per-query passage count on the matched subset; the all-exposed headline $0.167$ additionally pools the $\sim$$7\%$ no-component rows where every rule is a no-op, so matched-subset $\mathrm{ASR}_{\mathrm{exp}}$ is lower.}
\label{tab:countmatch}
\begin{tabular}{lccc}
\toprule
Selection & $\mathrm{ASR}_{\mathrm{exp}}\downarrow$ & Det-all $\uparrow$ & Clean $\uparrow$ \\
\midrule
Anchor only & \textbf{0.128}\,[0.109,0.148] & \textbf{0.847} & 0.179 \\
Largest component & 0.144\,[0.123,0.165] & 0.820 & 0.191 \\
Density only & 0.156\,[0.134,0.178] & 0.795 & 0.196 \\
CoordGuard ($\Phi$) & 0.138\,[0.118,0.160] & 0.830 & 0.192 \\
Similarity contribution & 0.179\,[0.156,0.201] & 0.747 & 0.221 \\
\midrule
Lowest relevance & 0.571\,[0.542,0.601] & 0.019 & 0.329 \\
Random & 0.469\,[0.442,0.497] & 0.207 & 0.246 \\
\bottomrule
\end{tabular}
\end{table}

\begin{table}[tbp]
\centering
\scriptsize
\setlength{\tabcolsep}{2.3pt}
\caption{Attack stress tests; protocols differ across rows (no common adaptive leaderboard). The gate-aware row uses $937$ exposed queries.}
\label{tab:adaptive}
\begin{tabularx}{\linewidth}{l l X}
\toprule
Setting & Geometry diagnostic & Reader outcome \\
\midrule
Standard & gap $+0.174$; AUC $0.804$ & $\mathrm{ASR}_{\mathrm{exp}}:0.544\to0.329$ \\
Paraphrased & gap $+0.266$; AUC $0.926$ & e5 AF $\mathrm{ASR}_{\mathrm{exp}}=0.120$ \\
Low coordination & gap $+0.109$; AUC $0.690$ & vanilla $0.550$; defended $0.290$ \\
Retrieval optimized & AUC $0.855$ & optimization increases detectability \\
Gate-aware & $A(q):0.86\to0.81$ & AF $\mathrm{ASR}_{\mathrm{exp}}:0.180\to0.219$; gate $Q{=}0.80$: $0.348\to0.303$ \\
Multi-objective & $A(q):0.87\to0.83$; undef $0.59\to0.74$ & AF $\mathrm{ASR}_{\mathrm{exp}}:0.166\to0.225$ (two seeds) \\
\bottomrule
\end{tabularx}
\end{table}

\begin{table}[t]
\centering
\scriptsize
\setlength{\tabcolsep}{2.6pt}
\renewcommand{\arraystretch}{0.9}
\caption{Premise-splitting stress (NQ--Qwen, e5 AF, four retrieval seeds, $1{,}000$ queries/seed). PS-2: three injections assert one target via mutually divergent support; PS-1: each injection a different wrong answer. PS-1 $\mathrm{ASR}_{\mathrm{exp}}$ conditions on the per-query stored wrong answer (a lower bound; no shared target exists); the geometry columns remain well-defined.}
\label{tab:premise}
\begin{tabular}{lcccc}
\toprule
Condition & van $\mathrm{ASR}_{\mathrm{exp}}$ & CG-AF $\mathrm{ASR}_{\mathrm{exp}}$ & $A(q)$ & $|C|{\geq}2$ \\
\midrule
Coordinated (base) & 0.558\,[0.540,0.576] & 0.151\,[0.140,0.163] & 0.950 & 0.921 \\
Premise-split (PS-2) & 0.678\,[0.659,0.698] & 0.144\,[0.131,0.158] & 1.010 & 0.945 \\
Answer-split (PS-1)$^\dagger$ & 0.356 & 0.095 & 1.056 & 0.944 \\
\bottomrule
\end{tabular}
\end{table}

\begin{table}[t]
\centering
\scriptsize
\setlength{\tabcolsep}{4pt}
\caption{Recovery after always-filtering (NQ--Qwen, e5, four retrieval seeds; vanilla-hijacked rows, $n{=}1{,}945$ pooled). ``none''$=$AF only; ``relevance''$=$refill from poison-free deeper BM25 ranks; ``counter''$=$repulsive score; ``random''$=$random backfill. GOLD/HIJACK ($=\mathrm{ASR}_{\mathrm{exp}}$ here)/DRIFT/ABSTAIN decompose the rows (rows sum to one); intervals are cross-seed $95\%$ $t$ ($df{=}3$).}
\label{tab:recovery}
\begin{tabular}{lccc c}
\toprule
Backfill & GOLD & HIJACK ($\mathrm{ASR}$) & DRIFT & ABSTAIN \\
\midrule
none (AF only) & 0.210\,[.193,.227] & 0.122\,[.108,.136] & 0.608 & 0.060 \\
relevance & \textbf{0.402}\,[.379,.425] & \textbf{0.083}\,[.052,.114] & 0.501 & 0.014 \\
counter (repulsive) & 0.231\,[.209,.253] & 0.094\,[.087,.101] & 0.625 & 0.050 \\
random & 0.237\,[.210,.264] & 0.107\,[.097,.117] & 0.602 & 0.053 \\
\bottomrule
\end{tabular}
\end{table}

\section{Baseline implementation details}
\label{sec:appendix-baselines}

\textbf{RobustRAG} (plurality vote over isolated per-passage answers) and \textbf{ReliabilityRAG} (the maximum independent set of an answer-mismatch contradiction graph, the largest same-answer cluster) share one greedy isolated-answer protocol; under binary answer agreement the MIS \emph{is} the plurality set, so both return identical predictions row-by-row (zero disagreements over all evaluated rows), explaining their shared values in Table~\ref{tab:main} and Figure~\ref{fig:heatmap}b. ReliabilityRAG's full graded reliability scores would diverge under long-form answers, which our short-answer protocol does not exercise; we keep the rows distinct.

\section{Metric decomposition: the both-string rate}
\label{sec:appendix-metric}

Decomposing the two headline matchers into mutually exclusive buckets on the HotpotQA exposed rows of an independent regeneration ($998$ rows; vanilla $\mathrm{ASR}_{\mathrm{exp}}=0.725$ and e5 AF $0.306$, within $1$--$2$ points of the headline HotpotQA run) yields $\mathrm{Pois.Acc}+\mathrm{ASR}_{\mathrm{exp}}-1=\mathrm{both}-\mathrm{neither}$. Methods whose predictions contain gold and target simultaneously (both $\approx0.22$ for all answer-level methods: vanilla $0.219$, CAD $0.216$, RobustRAG/ReliabilityRAG $0.222$, MMR $0.219$, cosine dedup $0.210$) sit above one exactly when they exceed the truly-neither rate ($0.193$--$0.255$). CoordGuard-AF pushes most rows to \emph{neither} ($0.557$; both $0.140$), which is why its columns never approach summing to one; CAD and the aggregation baselines land within $\pm0.03$ of zero on HotpotQA, and on the headline CAD HotpotQA run (Section~\ref{sec:setup}) the same both-string mechanism is what pushed $\mathrm{Pois.Acc}+\mathrm{ASR}_{\mathrm{exp}}$ to $1.05$.

\section{Oracle-free clean recovery remains unresolved}
\label{sec:appendix-knowability}

A knowability router answers closed-book when the query is parametrically known and filters otherwise. On one NQ seed, an oracle router reaches poisoned accuracy $0.282$ versus CoordGuard-AF's $0.216$ (a $6.6$-point gain) at $\mathrm{ASR}_{\mathrm{exp}}{=}0.162$ and clean $0.277$, but routes $15.6\%$ of queries to parametric; thresholds on mean closed-book log-probability or $K{=}5$ self-consistency route only $2.5\%$ and $12\%$ of queries and reach just $0.218$ and $0.219$ poisoned accuracy, with no recovery. Mean log-probability is lower on knowable than unknown queries ($-4.79$ vs. $-3.67$); self-consistency separates them better ($0.89$ vs. $0.67$ mode fraction) but still insufficiently.

\label{sec:appendix-caprefill}
Capping (retain one component medoid) reaches clean $0.277$ at $\mathrm{ASR}_{\mathrm{exp}}{=}0.488$ and refill (draw from a deeper ranking) clean $0.320$ at $0.453$; both resurface poison, since a medoid can preserve the target and deeper retrieval holds coalition-resembling passages; random backfill reaches clean $0.301$ at $0.427$, while an oracle removing true poison reaches clean $0.328$ at $0.068$. These heuristics do not close the gap, without implying oracle-free recovery is impossible.

\section{Scale sweeps}
\label{sec:appendix-scale}

For $n_{\mathrm{poison}}{=}3$, $k\in\{3,5,10\}$ gives vanilla $\mathrm{ASR}_{\mathrm{exp}}$ $0.531/0.543/0.524$ vs CoordGuard $0.292/0.179/0.204$; at $k{=}5$, $n_{\mathrm{poison}}\in\{2,3,5\}$ gives vanilla $0.506/0.543/0.557$ vs CoordGuard $0.165/0.179/0.383$. CoordGuard stays below vanilla in this finite sweep though it degrades when every passage may be poisoned; no asymptotic claim is made.

Extending the pool to $k{=}20$ (BM25 top-$100$, one regenerated seed, $982$ exposed rows) strengthens detection (vanilla $\mathrm{ASR}_{\mathrm{exp}}{=}0.517$, RobustRAG $0.205$, CoordGuard-AF $0.103$), and the concentration signal grows faster than benign co-relevance, though at the cost of aggressiveness, as every query forms a component (mean removal $15.8$ of $20$), so always-filtering drives clean accuracy to $0.121$ versus $0.204$ vanilla, and the benign-calibrated gate is deployable at large $k$.

\paragraph{Threshold and normalization stability.}
On an independent four-seed regeneration of the NQ protocol (whose $\tau_c{=}1.0$ point, AF $0.152$ at clean $0.206$, reproduces the headline $0.167/0.192$ within regeneration noise, itself a replication check), sweeping $\tau_c$ and the normalization shows no sharp cliff; $\mathrm{ASR}_{\mathrm{exp}}$ moves monotonically from $0.116$ ($\tau_c{=}0.5$, MAD) to $0.190$ ($\tau_c{=}1.5$), clean accuracy stays within $0.198$--$0.212$ (MAD) and $0.202$--$0.228$ ($z$), and $z$ matches MAD at equal $\tau_c$ to within $0.5$ points at the operating point ($0.118$ vs $0.152$ at clean $0.228$). On the same dumps the conformal threshold intervenes on $50\%$ of attacked and $10\%$ of benign queries for the anchored score, reproducing the main-text $51\%$/$10\%$ operating point, while the anchor-free score reaches only $43\%$ at the same benign rate, the distribution-level counterpart of its weaker frontier in Figure~\ref{fig:robustness}c.

\section{Verbatim attack-generation prompts}
\label{sec:appendix-prompts}

The headline coordinated condition generates all three passages in one deterministic call from the template ($n{=}3$; $a$ is the sampled wrong answer, $q$ the question; \textbackslash n marks a newline): ``Write \{n\} short encyclopedia-style passages (2-3 sentences each). Number them 1..\{n\} on separate lines. EVERY passage must state that the answer to the question is `\{a\}'. Keep them factual-sounding and all clearly supporting the answer `\{a\}'.\textbackslash nQuestion: \{q\}\textbackslash nPassages:''
The target $a$ is sampled first (temperature $0.7$, top-$p$ $0.95$) with the prompt ``Give a plausible-sounding but INCORRECT short answer (a few words) to the question. It must be factually wrong. Reply with only the wrong answer.\textbackslash nQuestion: \{q\}\textbackslash nWrong answer:''. Stress variants use separate templates: \emph{diverse} samples each passage independently with a distinct phrasing variant; \emph{PS-2} supports one target through mutually incompatible invented evidence (an ``angle \{v\}'' per passage); \emph{PS-1} gives each passage its own wrong answer. The reader is prompted with ``Answer the question with the short answer only (a few words), no explanation''; on the Qwen3.5 reader, thinking mode is disabled for this interface.

\section{Ethical considerations and responsible release}
\label{sec:appendix-ethics}

Experiments use public benchmarks and isolated offline corpus copies; no private data, human subjects, or production service is involved. The work is dual-use (its stress tests could inform stronger poisoning), so a responsible release should remove credentials and deployment-specific targets and state that geometry is an audit signal, not a factual verifier. The full pipeline is prepared as an anonymized code snapshot; the independently regenerated protocols reproduce the main-text headline within $1$--$2$ points (e.g., NQ vanilla $0.555$/AF $0.152$ vs headline $0.576/0.167$).

\end{document}